\documentclass[11pt,a4paper]{article}

\usepackage[margin=1in]{geometry}
\usepackage{amsmath}
\usepackage{amssymb}
\usepackage{booktabs}
\usepackage{multirow}
\usepackage{array}
\usepackage{cite}
\usepackage{placeins}
\usepackage{float}
\usepackage{tabularx}
\usepackage{graphicx}
\usepackage[colorlinks=true, allcolors=blue]{hyperref}

\newcolumntype{L}[1]{>{\raggedright\arraybackslash}p{#1}}

\begin{document}

\title{DKEKAN: A single-parameterized KAN surrogate for Drift Kinetic Equation Toward Fast Neoclassical Toroidal Viscosity Torque Modeling in Tokamaks}

\author{
Jinpeng Huang$^{1}$, Xingting Yan$^{2,*}$, Mingyu Zhang$^{1}$, Nana Bao$^{1,2}$,\\
Zixuan Song$^{1}$, Yuetao Meng$^{1}$, Weiyong Zhou$^{1}$ and Youwen Sun$^{2}$\\[0.5em]
\small $^{1}$ School of Internet, Anhui University, Hefei, Anhui 230039, China\\
\small $^{2}$ Hefei Institutes of Physical Science, Chinese Academy of Sciences, Hefei, Anhui 230031, China\\
\small $^{*}$Author to whom any correspondence should be addressed.\\
\small E-mail: xingting.yan@ipp.ac.cn
}

\date{}

\maketitle

\begin{abstract}
The neoclassical toroidal viscosity (NTV) torque is a critical driver of toroidal rotation in tokamaks, profoundly influencing plasma stability and performance. Consequently, incorporating NTV effects is essential for modern integrated modeling frameworks that aim to self-consistently unify multiple physical processes. However, the high computational cost of NTV modeling precludes its self-consistent integration within such frameworks. This bottleneck arises because NTV calculation requires solving its governing equation—the drift kinetic equation (DKE)—in high-dimensional phase space. To address this issue, this study develops DKEKAN, a single-parameterized Kolmogorov--Arnold Network (SKAN) surrogate for solving DKE, to realize fast NTV modeling in tokamaks. The research process consists of the following steps: Firstly, a large dataset mapping DKE equation parameters to solutions is generated based on first-principle simulations under plasma parameters of the Experimental Advanced Superconducting Tokamak (EAST); Secondly, a surrogate model for solving DKE is developed based on the SKAN framework, which also incorporates a modular expert network design; Finally, the DKEKAN surrogate model is integrated with the NTV modeling framework to realize fast NTV calculation. With its physics-grouped expert layer and SKAN backbone, DKEKAN outperforms the tested MLP, KAN, and neural-operator baselines in overall prediction accuracy, while reducing the standalone DKE-solving time from 35.85~s to 3.74~s, corresponding to a speedup of approximately \(9.6\times\), and reducing the total coupled NTVTOK runtime from 38.24~s to 5.58~s, corresponding to an overall speedup of approximately \(6.9\times\). This work effectively overcomes the computational bottleneck in NTV simulations, thus supporting further integrated modeling that incorporates NTV effects.
\end{abstract}


\section{Introduction}

Neoclassical toroidal viscosity (NTV) torque is an important mechanism of toroidal momentum transport in tokamak plasmas subject to three-dimensional magnetic perturbations~\cite{Shaing2003,Park2009,Shaing2015}. Such perturbations can be externally applied by resonant magnetic perturbation (RMP) coils for edge-localized mode (ELM) control, or can arise from intrinsic error fields and magnetohydrodynamic instabilities. By modifying toroidal rotation and rotation shear, NTV torque can affect momentum balance, plasma stability, and overall confinement performance~\cite{Zhu2006,Garofalo2008,Sun2010JET,Honda2015,Yang2019,Yan2023,Sheng2024,Li2025,Moyer2017,Sun2021EAST}. Reliable NTV modeling is therefore important for interpreting present tokamak experiments and for developing predictive integrated-modeling capabilities for future fusion devices.

In high-fidelity NTV models, the perturbed distribution function and the corresponding torque density are obtained by solving the drift kinetic equation (DKE). Several models have been developed to reduce the computational cost, including smoothly connected analytic formulae~\cite{Shaing2010,Liu2013} and approaches based on simplified collision operators, such as IPEC-PENT~\cite{Logan2013} and JOREK-PENTRC~\cite{Kim2022}. These models have enabled practical NTV calculations in many applications, but they may approximate or omit physical effects that are important in realistic regimes, including pitch-angle scattering, bounce-drift resonance, transitional collisionality, and realistic toroidal geometry effects. The NTVTOK code provides a more complete numerical treatment by solving the DKE with key neoclassical physics retained~\cite{Sun2010TEXTOR,Sun2019}, and has been used in a range of tokamak NTV studies~\cite{Sun2010JET,Sun2012JET,Li2019HL2A,Li2021EAST,Yan2021DIIID,Yan2023}. However, the DKE solver remains the dominant computational bottleneck in the NTVTOK workflow, limiting the routine incorporation of high-fidelity NTV physics into integrated or nonlinear plasma simulations.

Machine-learning surrogate modeling offers a promising route to accelerate expensive plasma-physics calculations. Recent work such as NTVTOK-ML has shown that data-driven models can substantially reduce the cost of NTV torque prediction~\cite{Yan2025NTVTOKML,Bao2026NTVML}. However, an end-to-end torque surrogate maps plasma parameters directly to the final torque profile and does not explicitly preserve the intermediate DKE solution. This makes it difficult to diagnose whether prediction errors originate from the perturbed distribution function or from the subsequent torque evaluation. Moreover, direct torque regression may become less robust when plasma parameters move outside the training distribution~\cite{Yan2025NTVTOKML}. In contrast, the present work focuses on a data-driven module-level surrogate trained from high-fidelity NTVTOK solutions and coupled back to the original torque-evaluation workflow. By replacing only the DKE-solving component, this strategy retains the remaining physics-based NTVTOK workflow, preserves physically meaningful intermediate quantities, and enables validation at both the DKE-solution and NTV-torque levels.

The DKE surrogate considered here is a data-driven parametric kinetic-equation learning problem with heterogeneous physical inputs. Neural operator methods, including Fourier Neural Operators (FNOs)~\cite{Li2021FNO} and DeepONets~\cite{Lu2021DeepONet}, provide natural baselines because they are designed to learn mappings between input functions and solution fields. In the present application, however, the surrogate must be repeatedly evaluated inside an existing NTV calculation workflow, where inference efficiency, memory footprint, and ease of coupling are important in addition to prediction accuracy~\cite{Wu2025TurboFNO,Karumuri2026PILNO}. Kolmogorov--Arnold Networks (KANs) have recently been proposed as a potential alternative to conventional multilayer perceptrons, replacing fixed nodal activation functions with learnable univariate functions on network edges~\cite{Liu2025KAN}. Lightweight variants such as SKAN have been proposed to improve the accuracy-efficiency tradeoff~\cite{Chen2024SKAN}, making them attractive for repeated inference inside a physics-based simulation workflow.

In this work, we develop DKEKAN, a structured neural surrogate for the DKE module in NTVTOK. Unlike end-to-end torque surrogates, DKEKAN replaces only the DKE-solving module while retaining the subsequent physics-based torque evaluation. The model encodes heterogeneous DKE inputs through deterministic expert pathways before feature fusion and uses a lightweight SKAN backbone to learn the nonlinear coefficient-to-solution mapping. We evaluate DKEKAN using discharge-level splitting, comparisons with MLP-, KAN-, and neural-operator-based baselines, and parameter-extrapolation tests beyond the original EAST training regime. The trained surrogate is then coupled back into NTVTOK to quantify the DKE-level accuracy, the fidelity of the resulting electron and ion NTV torque profiles, and the end-to-end acceleration of the coupled workflow.

The remainder of this paper is organized as follows. Section~\ref{sec:dke_ntvtok} introduces the DKE formulation and the computational cost of the DKE solver within the NTVTOK workflow. Section~\ref{sec:dataset} describes the EAST-based dataset generation, partitioning, and preprocessing. Section~\ref{sec:network_design} presents the DKEKAN architecture and the baseline models. Section~\ref{sec:experimental_setup} reports the surrogate accuracy, NTVTOK-coupled torque prediction, computational speedup, and extrapolation tests. Section~\ref{sec:conclusion} summarizes the main conclusions.

\section{Drift kinetic equation and NTVTOK workflow}
\label{sec:dke_ntvtok}

In NTVTOK, the NTV torque is calculated from the drift-kinetic response of trapped particles to non-axisymmetric magnetic perturbations. Such perturbations modify the magnetic-field strength along the field line and introduce a helical ripple. For trapped particles, this ripple produces a nonambipolar neoclassical particle flux. The ambipolarity-restoring response associated with this flux can be expressed as an effective toroidal torque~\cite{Sun2019}.

The perturbed distribution function is the central quantity in this calculation. After this distribution function is obtained, NTVTOK performs the pitch-angle and energy integrations and evaluates the electron and ion torque-density profiles. In the present work, the surrogate model is introduced at the DKE-solving step. The downstream torque-evaluation procedure in NTVTOK is kept unchanged, so that the surrogate replaces the repeated numerical solution of the kinetic equation rather than the physical torque model itself.

\subsection{Drift kinetic equation for NTV torque calculation}
\label{subsec:dke_formulation}

The NTVTOK formulation retains bounce-drift resonance effects by introducing a coordinate transformation associated with the bounce and drift motion of trapped particles. The linearized drift kinetic equation is then Fourier decomposed in the transformed coordinates~\cite{Sun2019}. For a given toroidal harmonic number \(n\) and bounce harmonic number \(l\), the normalized equation can be written as
\begin{equation}
    I_1 L_0\left(\hat{f}_{nl}\right)
    - i I_2 \hat{f}_{nl}
    - i \hat{b}_{nl}
    = 0 ,
    \label{eq:harmonic_dke}
\end{equation}
where \(L_0\) is the pitch-angle scattering collision operator. The coefficient \(I_1\) represents the normalized collisionality:
\begin{equation}
    I_1 =
    \frac{\nu_d/(2\epsilon)}{I_0},
    \qquad
    I_0 =
    \sqrt{
    \left[
    \frac{\nu_d}{2\epsilon}
    \right]^2
    +
    \max\left(\Omega_{nl}^2\right)
    },
    \label{eq:I1_definition}
\end{equation}
where \(\nu_d\) is the deflection frequency and \(\epsilon\) is the inverse aspect ratio. The coefficient \(I_2\) contains the drift and bounce frequencies:
\begin{equation}
    I_2 =
    -\frac{\Omega_{nl}}{I_0},
    \qquad
    \Omega_{nl}
    =
    n\left(\Omega_E+\Omega_B\right)
    +
    l\omega_b .
    \label{eq:I2_definition}
\end{equation}
Here \(\Omega_E\) is the \(\boldsymbol{E}\times\boldsymbol{B}\) drift frequency, \(\Omega_B\) is the magnetic drift frequency, and \(\omega_b\) is the bounce frequency. These frequencies are evaluated in NTVTOK from the magnetic equilibrium and kinetic profiles~\cite{Sun2019}. The perturbation enters through \(\hat{b}_{nl}\), the normalized Fourier component of the non-axisymmetric magnetic-field strength. The unknown \(\hat{f}_{nl}\) is the normalized perturbed distribution function for the corresponding harmonic.

The pitch-angle variable is defined as
\begin{equation}
    \kappa^2 =
    \frac{v^2/2-\mu B_m}
    {\mu\left(B_M-B_m\right)}
    \in [0,1],
    \label{eq:kappa_definition}
\end{equation}
where \(v\) is the particle speed, \(\mu\) is the magnetic moment, and \(B_m\) and \(B_M\) are the minimum and maximum magnetic-field strengths on a flux surface. In this coordinate, the pitch-angle scattering operator contains derivative terms with respect to \(\kappa^2\). Using the notation adopted in the present surrogate dataset, it is written as
\begin{equation}
    L_0\left(\hat{f}_{nl}\right)
    =
    F_1 \frac{\partial \hat{f}_{nl}}{\partial \kappa^2}
    +
    F_2
    \left[
    \frac{\partial^2 \hat{f}_{nl}}{\partial (\kappa^2)^2}
    -
    \left(\frac{\partial D}{\partial \kappa^2}\right)^2
    \hat{f}_{nl}
    \right],
    \label{eq:collision_operator}
\end{equation}
where \(F_1\) and \(F_2\) are coefficient profiles from the collision operator, and \(D\) comes from the coordinate transform. The detailed expressions of these quantities follow the NTVTOK model~\cite{Sun2019}. Once \(\hat{f}_{nl}\) is solved, it is passed to the subsequent pitch-angle and energy integrations for the nonambipolar particle fluxes and the NTV torque density.

To keep the module interface in NTVTOK unchanged, we introduce the following coefficient notation for the DKE surrogate:
\begin{equation}
    C_1 = I_1, \qquad
    C_2 =
    - I_1 F_2
    \left(
    \frac{\partial D}{\partial \kappa^2}
    \right)^2
    - i I_2, \qquad
    C_3 = -i \hat{b}_{nl}, \qquad
    F = \hat{f}_{nl}.
    \label{eq:coefficient_definition}
\end{equation}

The numerical DKE solver may then be viewed as the operator
\begin{equation}
    \mathcal{G}:
    \left\{
    F_1,F_2,C_1,C_2,C_3
    \right\}
    \longmapsto F .
    \label{eq:operator_mapping_complex}
\end{equation}

The quantities \(C_2\), \(C_3\), and \(F\) are complex-valued. In the neural-network implementation, their real and imaginary parts are treated separately,
\begin{equation}
    C_2 = C_{2,r} + i C_{2,i}, \qquad
    C_3 = C_{3,r} + i C_{3,i}, \qquad
    F = F_r + i F_i .
    \label{eq:real_imag_decomposition}
\end{equation}
This real-valued representation is adopted when the complex DKE is used as a surrogate-learning target. The learned mapping is therefore
\begin{equation}
    \mathcal{G}_{\theta}:
    \left\{
    F_1,F_2,C_1,C_{2,r},C_{2,i},C_{3,r},C_{3,i}
    \right\}
    \longmapsto
    \left\{
    F_r,F_i
    \right\},
    \label{eq:operator_mapping_real}
\end{equation}
where \(\theta\) denotes the trainable parameters of the surrogate.

The input variables in Eq.~\eqref{eq:operator_mapping_real} have different mathematical forms and physical roles. The quantities \(F_1\), \(F_2\), \(C_2\), and \(C_3\) are profiles on the \(\kappa^2\) grid, whereas \(C_1\) is a scalar. 

\subsection{Computational cost of the DKE solver}
\label{subsec:module_acceleration}

A standard NTVTOK calculation first constructs the DKE coefficients from the magnetic equilibrium, kinetic profiles, and non-axisymmetric perturbation fields. The DKE is then solved on the required phase-space grid and Fourier harmonics. The resulting perturbed distribution function is finally passed to the torque-evaluation part of the code to compute the electron and ion NTV torque-density profiles.

\begin{figure}[H]
\centering
\includegraphics[width=0.7\textwidth]{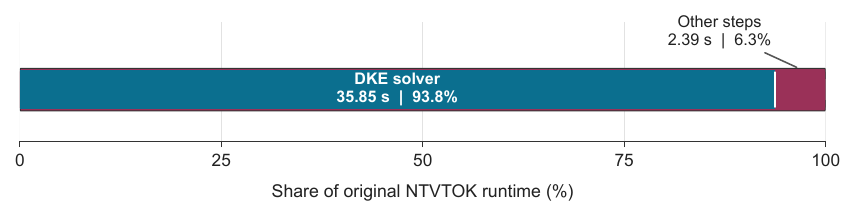}
\caption[Runtime decomposition of the original NTVTOK workflow]{
    Runtime decomposition of the original NTVTOK workflow for NTV torque evaluation.
    The DKE solver requires 35.85~s, accounting for 93.8\% of the total NTVTOK runtime of 38.24~s, while the remaining workflow contributes 2.39~s, or 6.3\%.
    This identifies the DKE solver as the dominant computational bottleneck in the original NTVTOK calculation.
}
\label{fig:dke_times}
\end{figure}

In this workflow, the numerical solution of the DKE is the main computational bottleneck, as indicated by the runtime comparison in Fig.~\ref{fig:dke_times}. For a small number of cases, this cost is still manageable. It becomes restrictive, however, when NTVTOK is used for parameter scans or integrated modeling, where the torque calculation has to be repeated over many plasma conditions. The limitation is even more direct in nonlinear simulations, since the plasma state evolves in time and the NTV torque may need to be updated repeatedly. In such applications, repeatedly recomputing the full DKE solution can dominate the cost of the NTV module.

\FloatBarrier

\section{Dataset generation and preprocessing}
\label{sec:dataset}

To construct the dataset for DKE surrogate training, 12 EAST discharges associated with edge-localized mode (ELM) control experiments were selected as the baseline discharge set. The edge safety factor \(q_{95}\) of these discharges lies approximately in the range 3.7--5.3. This range covers typical operating conditions of RMP-assisted ELM control experiments on EAST, where the plasma response to resonant magnetic perturbations is sensitive to the safety factor. The selected discharges give an experimentally relevant parameter space for training and testing the DKE surrogate.

Before machine-learning samples were generated, the raw discharge data were processed through the NTVTOK workflow. The magnetic equilibrium, kinetic profiles, and non-axisymmetric perturbation information were assembled and converted into DKE coefficients. In the present surrogate study, the dataset is restricted to the selected harmonic component with \(n=1\) and \(l=0\) to illustrate the key ideas used in the model development. The dataset contains separate electron and ion cases, corresponding to the electron and ion contributions to the NTV torque calculation. 
For each baseline discharge and particle species, the DKE parameters over the phase-space grid and relevant Fourier harmonics produce 4,103,125 candidate samples. Across the 12 discharges, this yields 49,237,500 candidate samples per species.

To avoid an overly optimistic estimate of generalization from random sample-level splitting, a discharge-level split was used. The 12 discharges were sorted according to the plasma current \(I_p\). The 10 lower-\(I_p\) discharges were used for training, while the remaining two higher-\(I_p\) discharges were reserved for validation and testing, respectively. The split separates the training and test data at the discharge level and introduces a shift from lower-current to higher-current operation. Higher \(I_p\) discharges are generally associated with higher performance operations, which are more relevant to future reactor-level plasma regimes. This setting gives a more stringent test of model generalization than a purely random sample split.

Based on this discharge-level partition, \(8\times10^4\) samples were randomly drawn from each discharge for model training, validation-based model selection, and final evaluation. This resulted in \(8\times10^5\) samples for the training set, \(8\times10^4\) samples for the validation set, and \(8\times10^4\) samples for the test set. The effect of sample size on feature-space coverage was also examined. The statistical distribution and coverage of the input features became nearly saturated when the total number of samples approached \(10^6\). Further increasing the sample size did not introduce appreciable new regions in the feature space. Approximately \(10^6\) samples were used for training, validation, and testing in each particle case.

The model inputs are the DKE coefficients defined in Sec.~\ref{sec:dke_ntvtok},
\[
\{F_1,F_2,C_1,C_{2,r},C_{2,i},C_{3,r},C_{3,i}\},
\]
whose statistical ranges are listed in Table~\ref{tab:features}. Here, \(F_1\), \(F_2\), \(C_{2,r}\), \(C_{2,i}\), \(C_{3,r}\), and \(C_{3,i}\) are 200-dimensional profiles on the \(\kappa^2\) grid, while \(C_1\) is a scalar input. The model output is the DKE solution \(F\), represented by its real and imaginary components, \(F_r\) and \(F_i\), each with dimension 200.

\begin{table*}[t]
\centering
\footnotesize
\setlength{\tabcolsep}{3pt}
\renewcommand{\arraystretch}{1.12}
\caption{Statistical ranges and mean values of the DKE input features in the sampled electron and ion datasets.}
\label{tab:features}
\begin{tabular}{cccccc}
\toprule
Input feature & Dimension & Electron range & Electron mean & Ion range & Ion mean \\
\midrule
$F_1$ & 200 & $[1.0000, 2.1175]$ & $1.2756$ & $[1.0000, 2.1175]$ & $1.2752$ \\
$F_2$ & 200 & $[9.521\times 10^{-4}, 1.2699]$ & $0.5200$ & $[9.521\times 10^{-4}, 1.2699]$ & $0.5200$ \\
$C_{1}$ & 1 & $[0.0022, 1.0000]$ & $0.7762$ & $[2.7\times 10^{-6}, 1.0000]$ & $0.1741$ \\
$C_{2,r}$ & 200 & $[-7290, -7.076\times 10^{-16}]$ & $-0.3083$ & $[-4.034\times 10^{6}, -1.169\times 10^{-12}]$ & $-94.3549$ \\
$C_{2,i}$ & 200 & $[-1.0000, 1.0000]$ & $-0.1206$ & $[-1.0000, 1.0000]$ & $0.0688$ \\
$C_{3,r}$ & 200 & $[-1.0000, 1.0000]$ & $-0.0175$ & $[-1.0000, 1.0000]$ & $-0.0185$ \\
$C_{3,i}$ & 200 & $[-1.0000, 0.9552]$ & $-0.0128$ & $[-1.0000, 0.9552]$ & $-0.0127$ \\
\bottomrule
\end{tabular}
\renewcommand{\arraystretch}{1.0}
\end{table*}

Among the input variables, \(C_{2,r}\) has the largest dynamic range, spanning several orders of magnitude. Directly using this quantity as a network input would give this channel a much larger numerical scale than the other inputs and could disturb the relative weighting among physical variables during training. A signed logarithmic transformation was applied to \(C_{2,r}\):
\begin{align}
C_{2,r,e}^{\log} &= \operatorname{sgn}(C_{2,r,e})
\log_{10}\left(C_e|C_{2,r,e}|+1\right), \\
C_{2,r,i}^{\log} &= \operatorname{sgn}(C_{2,r,i})
\log_{10}\left(C_i|C_{2,r,i}|+1\right),
\end{align}
where \(\operatorname{sgn}(\cdot)\) is the sign function; \(C_e\) and \(C_i\) are the scaling constants for the electron and ion cases, respectively; and \(C_{2,r,e}\) and \(C_{2,r,i}\) denote the values of \(C_{2,r}\) in the electron and ion datasets.

The scaling constants were set to \(C_e=C_i=10^6\). This value was selected after scanning \(C\) over the range \(10^2\)--\(10^{10}\). Over this interval, the model accuracy showed only a weak dependence on \(C\), with the maximum variation in \(R^2\) being approximately 2\%. A common scaling constant was used for the electron and ion cases to keep the preprocessing procedure consistent.

\FloatBarrier

\section{Network Design}
\label{sec:network_design}

\subsection{Baseline models}
\label{subsec:baseline_models}

A multilayer perceptron (MLP) is used as the basic reference model for the DKE surrogate task. The MLP adopts a standard fully connected architecture with ReLU activations. Although simple, this baseline is useful for assessing how far the DKE mapping can be represented by a conventional finite-dimensional nonlinear regressor.

Two operator-learning models, Fourier Neural Operator (FNO) and DeepONet, are also included for comparison. The DKE surrogate problem has a natural operator-learning interpretation: the input coefficients specify a parameterized kinetic equation, while the output is the corresponding solution evaluated on the \(\kappa^2\) grid. This makes neural-operator models relevant baselines, particularly for judging whether a more specialized architecture is needed for the present DKE data.

FNO learns mappings between function spaces by parameterizing integral operators in Fourier space~\cite{Li2021FNO}. In a typical FNO layer, the input field is transformed to spectral space, a truncated set of Fourier modes is updated by learnable weights, and the result is transformed back to physical space. This spectral construction is well suited to problems in which the solution depends on spatially distributed input functions. In the present DKE task, the profile-like inputs \(F_1\), \(F_2\), \(C_2\), and \(C_3\) make FNO a meaningful operator-learning baseline.

DeepONet provides a different operator-learning formulation based on a branch--trunk decomposition~\cite{Lu2021DeepONet}. The branch network encodes the input functions or equation parameters, while the trunk network represents the coordinates at which the output function is evaluated. Their combined representation approximates the nonlinear operator from the input coefficients to the solution field. In this work, DeepONet is used to learn the mapping from DKE coefficients to the real and imaginary parts of the perturbed distribution function.

\subsection{KAN and SKAN}
\label{subsec:kan_skan}

We further consider the Kolmogorov--Arnold Network (KAN) framework as the basis for the backbone design~\cite{Liu2025KAN}. Motivated by the Kolmogorov--Arnold representation theorem, KAN replaces the fixed nodal activation functions used in conventional MLPs with learnable univariate functions placed on network edges. An \(L\)-layer KAN can be written as a composition of function matrices,
\begin{equation}
    f(x) = (\Phi_{L-1} \circ \Phi_{L-2} \circ \ldots \circ \Phi_0)(x),
    \label{eq:kan_composition}
\end{equation}
where \(\Phi_l\) denotes the function matrix at layer \(l\), and each element \(\phi_{l,j,i}: \mathbb{R} \rightarrow \mathbb{R}\) is a learnable nonlinear mapping from one input component to a neuron in the next layer.

In the original KAN, the edge functions \(\phi_{l,j,i}\) are represented by B-spline bases. For clarity, we refer to this original spline-based formulation as B-spline KAN hereafter. This representation gives KAN considerable flexibility, but it also introduces knot parameters and interpolation operations that increase the cost of forward evaluation compared with a standard MLP~\cite{Liu2025KAN}. Since the DKE surrogate is intended for repeated calls inside an NTV calculation workflow, inference cost is not a secondary issue. We therefore adopt the single-parameterized KAN (SKAN) design, following the Efficient KAN Expansion Hypothesis~\cite{Chen2024SKAN}. In particular, the learnable arctangent function, denoted as LArctan~\cite{Chen2024SKAN}, is used as the edge function:
\begin{equation}
    \mathrm{LArctan:}\quad \phi(x) = \arctan(kx),
    \label{eq:larctan}
\end{equation}
where \(k\) is the only trainable parameter. This form keeps the nonlinear edge representation while avoiding the heavier spline-based implementation, which is advantageous for a surrogate that must be evaluated many times in DKE and NTV calculations.

\subsection{DKEKAN architecture}
\label{subsec:dkekan_architecture}

The input variables of the DKE surrogate exhibit strong heterogeneity in terms of physical meaning, dimensional structure, and numerical scale. Specifically, \(C_1\) is a low-dimensional yet physically important coefficient, whereas \(F_1\), \(F_2\), \(C_2\), and \(C_3\) are high-dimensional profile-like physical quantities. A conventional concatenation-based architecture implicitly treats these variables as components of a single homogeneous feature vector and applies a uniform front-end mapping to all input groups. However, in the DKE solving task, different variable groups correspond to distinct physical mechanisms and differ substantially in both dimensionality and scale. Therefore, such a uniform processing scheme may not provide the most appropriate inductive bias for modeling heterogeneous physical inputs.

\begin{figure}[H]
\centering
\includegraphics[width=0.95\textwidth]{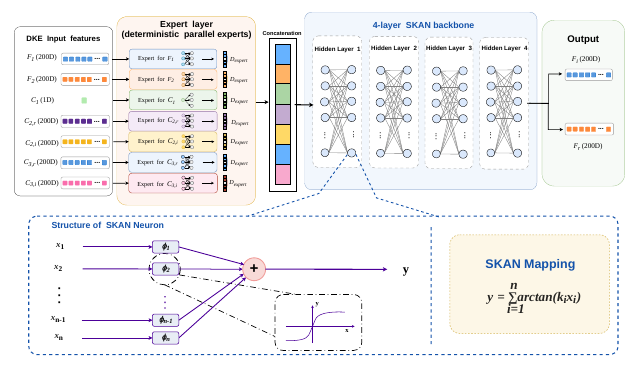}
\caption{Architecture of the proposed DKEKAN model. The heterogeneous DKE input groups, including $F_1$, $F_2$, $C_1$, and the real and imaginary components of $C_2$ and $C_3$, are first processed by deterministic group-wise expert pathways. The resulting latent features are concatenated and then passed to a four-layer SKAN backbone to predict the perturbed distribution functions. The lower panel illustrates the structure of a SKAN neuron and its arctangent-based nonlinear mapping used in the backbone.}
\label{fig:dkekan_architecture}
\end{figure}

A related feature-level modeling perspective can be found in Neural Additive Models~\cite{Agarwal2021NAM}, where individual input features are modeled through separate subnetworks to capture their nonlinear effects. This idea is consistent with our motivation that heterogeneous inputs should not necessarily be forced into a single uniform preprocessing pathway. In the present DKE surrogate task, this consideration is further supported by the physical grouping structure of the input variables. Therefore, this work proposes a new DKEKAN architecture, as illustrated in Fig.~\ref{fig:dkekan_architecture}, which incorporates the physical grouping prior of DKE variables and introduces a deterministic modular input interface before the SKAN backbone.

Specifically, let \(X_g\) denote the \(g\)-th physical input group, where
\begin{equation}
g \in \{F_1, F_2, C_1, C_{2,r}, C_{2,i}, C_{3,r}, C_{3,i}\}.
\end{equation}
Each physical variable group is mapped by a corresponding single-layer expert pathway \(E_g(\cdot)\) into a latent representation \(h_g\) with dimension \(D_{\mathrm{expert}}\):
\begin{equation}
h_g = E_g(X_g).
\end{equation}
The resulting group-wise latent representations are then concatenated and used as the input to the subsequent SKAN backbone:
\begin{equation}
h = \operatorname{Concat}
\left(
h_{F_1},
h_{F_2},
h_{C_1},
h_{C_{2,r}},
h_{C_{2,i}},
h_{C_{3,r}},
h_{C_{3,i}}
\right).
\end{equation}
The final prediction of the perturbed distribution function is obtained through the SKAN backbone:
\begin{equation}
\hat{F} = \operatorname{SKAN}(h).
\end{equation}

The assignment of the modular pathways is deterministic and follows the physical grouping structure of the DKE input variables. In this way, each physical variable group is separately embedded into a comparable latent space before fusion, enabling the subsequent SKAN backbone to model nonlinear interactions among physically meaningful latent representations.

\section{Experiments}
\label{sec:experimental_setup}

\subsection{Experimental Setup}

To provide a comprehensive evaluation of the proposed DKEKAN framework, eight neural surrogate models are constructed for comparison, including DKEKAN, SKAN, Expert MLP, MLP, Expert B-spline KAN, B-spline KAN, FNO, and DeepONet. Among these models, DKEKAN, Expert MLP, and Expert B-spline KAN share the same physics-grouped modular input interface, where heterogeneous DKE variables are first processed by group-specific expert branches and then concatenated before being passed to the corresponding backbone network. For the corresponding non-expert baselines, an equivalent first representation layer is introduced to ensure a controlled comparison. Specifically, the first-layer dimensions of MLP and SKAN are set to $7D_{\mathrm{expert}}$, matching the concatenated representation generated by the seven expert branches. For B-spline KAN, the first-layer dimension is set to 49, which is identical to the concatenated expert representation of Expert B-spline KAN when $D_{\mathrm{expert}}=7$. The architecture-specific search spaces and model-dependent settings are summarized in Table~\ref{tab:model_hyperparams}.

\begin{table}[!htbp]
\centering
\small
\renewcommand{\arraystretch}{1.2}
\caption{Unified Training Configuration}
\label{tab:training_config}
\makebox[\textwidth]{%
\begin{tabular}{c c}
\toprule
Item & Setting \\
\midrule
Optimizer & AdamW \\
Learning rate & $10^{-3}, 10^{-4}$ \\
Maximum epochs & 1000 \\
Checkpoint selection & Best validation loss \\
Scheduler & ReduceLROnPlateau \\
Scheduler factor & 0.6 \\
Scheduler patience & 20 epochs \\
Minimum learning rate & $10^{-6}$ \\
Early stopping patience & 60 epochs \\
Loss function & SmoothL1Loss($\beta=1.0$) \\
\bottomrule
\end{tabular}}
\end{table}

All models are trained using backpropagation with the AdamW optimizer. The maximum number of training epochs is fixed at 1000. The learning rate is selected from $\{10^{-3}, 10^{-4}\}$, and the best checkpoint is chosen according to the validation loss. Unless otherwise specified, SmoothL1 loss with $\beta=1.0$ is used as the training objective. During training, the full validation set is evaluated for learning-rate scheduling and checkpoint selection, while the test set is reserved exclusively for final evaluation. The unified training protocol is reported in Table~\ref{tab:training_config}.

\begin{table*}[!htbp]
  \centering
  \small
  \setlength{\tabcolsep}{5pt}
  \renewcommand{\arraystretch}{1.13}
  \caption{Architecture-specific hyperparameter search space for the neural surrogate models.}
  \label{tab:model_hyperparams}

  \begin{tabular*}{0.88\textwidth}{@{\extracolsep{\fill}} c c c @{}}
    \toprule
    \multicolumn{3}{c}{(a) Input representation and width search space} \\
    \midrule
    Model & Input representation & Width search space \\
    \midrule
    DKEKAN & Expert, $D_{\mathrm{expert}}\in\{20,50\}$ & SKAN width $\in\{200,300\}$ \\
    SKAN & Direct, first width $=7D_{\mathrm{expert}}$ & Hidden width $\in\{200,300\}$ \\
    Expert MLP & Expert, $D_{\mathrm{expert}}\in\{20,50\}$ & MLP width $\in\{200,300\}$ \\
    MLP & Direct, first width $=7D_{\mathrm{expert}}$ & Hidden width $\in\{200,300\}$ \\
    Expert B-spline KAN & Expert, $D_{\mathrm{expert}}=7$ & KAN width $\in\{40,48\}$ \\
    B-spline KAN & Direct, first width $=49$ & KAN width $\in\{40,48\}$ \\
    FNO & Direct operator input; head width $\in\{56,64\}$ & Channel width $\in\{12,14\}$ \\
    DeepONet & Branch--trunk input; first width $\in\{140,350\}$ & Hidden width $\in\{200,300\}$ \\
    \bottomrule
  \end{tabular*}

  \vspace{5pt}

  \begin{tabular*}{0.88\textwidth}{@{\extracolsep{\fill}} c c @{}}
    \toprule
    \multicolumn{2}{c}{(b) Depth and architecture-specific parameters} \\
    \midrule
    Model & Depth and architecture-specific parameters \\
    \midrule
    DKEKAN & 4 SKAN-style backbone layers \\
    SKAN & 4 SKAN-style layers \\
    Expert MLP & 4 fully connected backbone layers \\
    MLP & 4 fully connected layers \\
    Expert B-spline KAN & 4 B-spline KAN layers; grid size $\in\{5,7\}$; spline order $=3$ \\
    B-spline KAN & 4 B-spline KAN layers; grid size $\in\{5,7\}$; spline order $=3$ \\
    FNO & 3 FNO blocks; Fourier modes $\in\{10,12\}$ coupled with channel width \\
    DeepONet & 3 branch and trunk layers \\
    \bottomrule
  \end{tabular*}

  \renewcommand{\arraystretch}{1.0}
\end{table*}

All neural network models are implemented in PyTorch and trained on NVIDIA GeForce RTX 4090 GPUs with CUDA acceleration. After training, the selected neural surrogate models are integrated into the NTVTOK workflow for downstream physical simulation and validation in MATLAB R2023b.

\subsection{Model Comparison}
\label{subsec:model_comparison}

To assess the performance of DKEKAN, we compare it with seven baseline models: SKAN, Expert MLP, MLP, Expert B-spline KAN, B-spline KAN, DeepONet, and FNO. The comparison is not limited to the fitting accuracy of the DKE solution. Since the final goal is to replace the original DKE solver in the NTVTOK workflow, the runtime after coupling the surrogate models to the MATLAB-based physical solver is also included. The formal test results and the single-shot NTV torque evaluation time are summarized in Table~\ref{tab:formal_test_accuracy_runtime}.

\begin{table*}[!htbp]
  \centering
  \small
  \setlength{\tabcolsep}{5pt}
  \renewcommand{\arraystretch}{1.13}
  \caption{Formal test accuracy and single-shot NTV torque evaluation time of the best-selected surrogate models. The best value in each metric is highlighted in bold. RelMean denotes the mean relative error on the test set, and the runtime corresponds to one NTV torque evaluation after model integration into the physical solver.}
  \label{tab:formal_test_accuracy_runtime}

  \begin{tabular}{@{} c c c c c c @{}}
    \toprule
    Model & MSE  & MAE  & $R^2$  & RelMean  & Time (s) \\
    \midrule
    \multicolumn{6}{@{}c@{}}{(a) Ion test set} \\
    \midrule

    DKEKAN
    & $\mathbf{6.328{\times}10^{-2}}$
    & $\mathbf{4.043{\times}10^{-2}}$
    & $\mathbf{0.9436}$
    & $\mathbf{5.269{\times}10^{-2}}$
    & $3.72$ \\

    SKAN
    & $1.149{\times}10^{-1}$
    & $7.271{\times}10^{-2}$
    & $0.8977$
    & $8.176{\times}10^{-2}$
    & $4.13$ \\

    Expert MLP
    & $9.304{\times}10^{-2}$
    & $4.734{\times}10^{-2}$
    & $0.9172$
    & $5.854{\times}10^{-2}$
    & $\mathbf{2.44}$ \\

    MLP
    & $1.469{\times}10^{-1}$
    & $6.599{\times}10^{-2}$
    & $0.8692$
    & $7.501{\times}10^{-2}$
    & $3.01$ \\

    Expert B-spline KAN
    & $1.378{\times}10^{-1}$
    & $7.622{\times}10^{-2}$
    & $0.8773$
    & $1.061{\times}10^{-1}$
    & $21.69$ \\

    B-spline KAN
    & $1.695{\times}10^{-1}$
    & $9.024{\times}10^{-2}$
    & $0.8491$
    & $1.192{\times}10^{-1}$
    & $22.51$ \\

    FNO
    & $8.115{\times}10^{-2}$
    & $7.044{\times}10^{-2}$
    & $0.9277$
    & $1.230{\times}10^{-1}$
    & $7.10$ \\

    DeepONet
    & $1.311{\times}10^{-1}$
    & $6.855{\times}10^{-2}$
    & $0.8832$
    & $7.846{\times}10^{-2}$
    & $20.94$ \\

    \addlinespace[4pt]
    \midrule
    \multicolumn{6}{@{}c@{}}{(b) Electron test set} \\
    \midrule

    DKEKAN
    & $\mathbf{8.546{\times}10^{-5}}$
    & $\mathbf{3.849{\times}10^{-3}}$
    & $\mathbf{0.9988}$
    & $\mathbf{2.275{\times}10^{-2}}$
    & $3.72$ \\

    SKAN
    & $2.933{\times}10^{-4}$
    & $7.403{\times}10^{-3}$
    & $0.9958$
    & $3.005{\times}10^{-2}$
    & $4.13$ \\

    Expert MLP
    & $1.926{\times}10^{-4}$
    & $6.201{\times}10^{-3}$
    & $0.9972$
    & $3.251{\times}10^{-2}$
    & $\mathbf{2.44}$ \\

    MLP
    & $1.412{\times}10^{-3}$
    & $1.439{\times}10^{-2}$
    & $0.9796$
    & $7.135{\times}10^{-2}$
    & $3.01$ \\

    Expert B-spline KAN
    & $2.093{\times}10^{-4}$
    & $5.513{\times}10^{-3}$
    & $0.9970$
    & $3.003{\times}10^{-2}$
    & $21.69$ \\

    B-spline KAN
    & $1.903{\times}10^{-3}$
    & $1.615{\times}10^{-2}$
    & $0.9725$
    & $8.180{\times}10^{-2}$
    & $22.51$ \\

    FNO
    & $2.140{\times}10^{-4}$
    & $6.071{\times}10^{-3}$
    & $0.9969$
    & $3.491{\times}10^{-2}$
    & $7.10$ \\

    DeepONet
    & $6.846{\times}10^{-4}$
    & $9.865{\times}10^{-3}$
    & $0.9901$
    & $4.439{\times}10^{-2}$
    & $20.94$ \\

    \bottomrule
  \end{tabular}

  \renewcommand{\arraystretch}{1.0}
\end{table*}

\begin{figure}[H]
    \centering
    \includegraphics[width=0.95\textwidth]{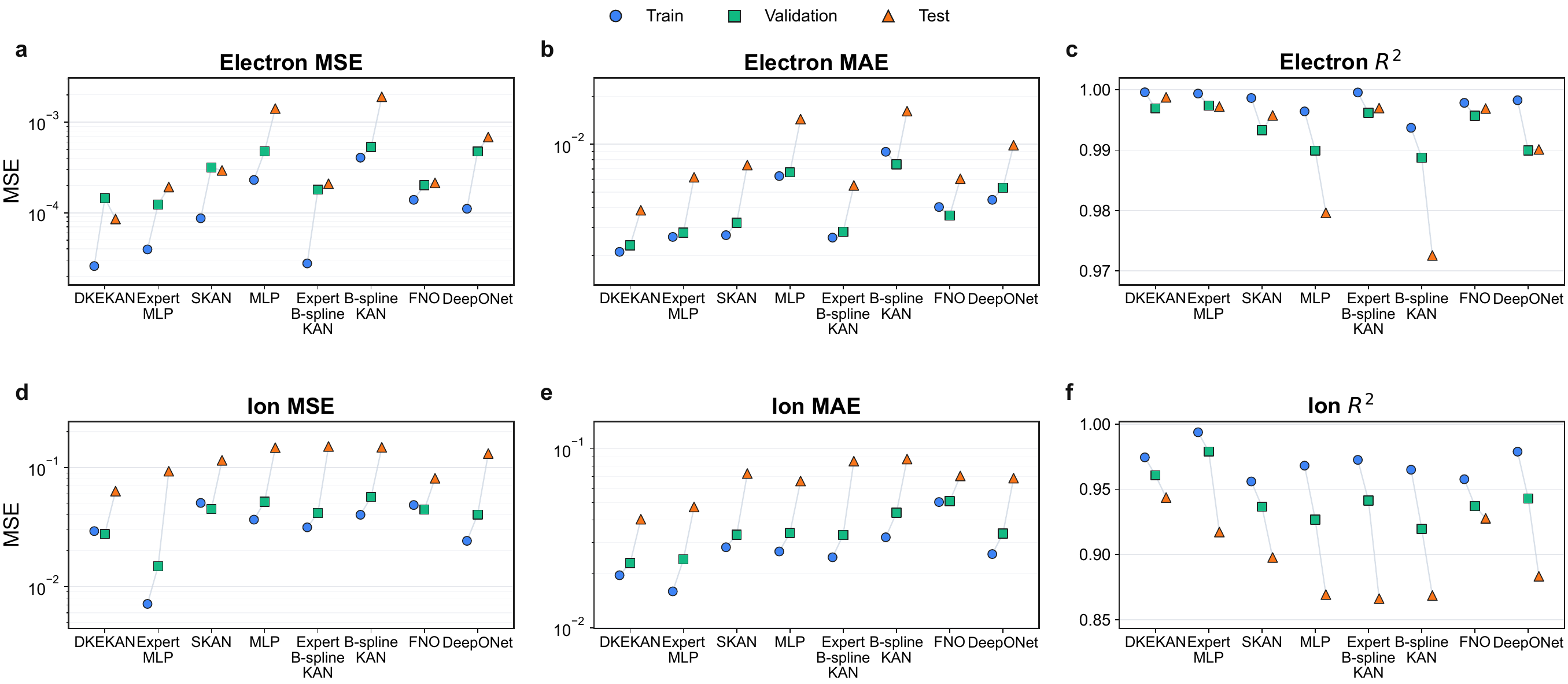}
    \caption{Predictive accuracy of the validation-selected best surrogate models. The figure compares the training, validation, and formal full-test performance of each model for both ion and electron cases. MSE and MAE are plotted on logarithmic axes, while $R^2$ is shown on a linear axis with a narrowed range to highlight the differences among models.}
    \label{fig:best_selected_accuracy}
\end{figure}

The prediction accuracy is evaluated by the mean squared error (MSE), mean absolute error (MAE), coefficient of determination ($R^2$), and mean relative error (RelMean). Fig.~\ref{fig:best_selected_accuracy} shows the training, validation, and formal full-test performance of the validation-selected best configuration of each model. The results are consistent with the values reported in Table~\ref{tab:formal_test_accuracy_runtime}. Among the compared models, DKEKAN gives the best overall test performance for both ion and electron cases. On the ion test set, it obtains the lowest MSE, MAE, and RelMean, together with the highest $R^2$. The same trend is observed for the electron test set, where DKEKAN again gives the most accurate prediction results.

The performance difference between the expert-enhanced models and their corresponding non-expert versions is also worth noting. Expert MLP improves over MLP, and Expert B-spline KAN performs better than B-spline KAN. Similarly, DKEKAN further improves the accuracy compared with SKAN. This comparison suggests that the expert mechanism is helpful for the present DKE-solving task. A possible reason is that the DKE solution contains nonlinear and multi-scale features, and these features are not always well represented by a single global mapping.

\begin{figure}[H]
    \centering
    \includegraphics[width=0.95\textwidth]{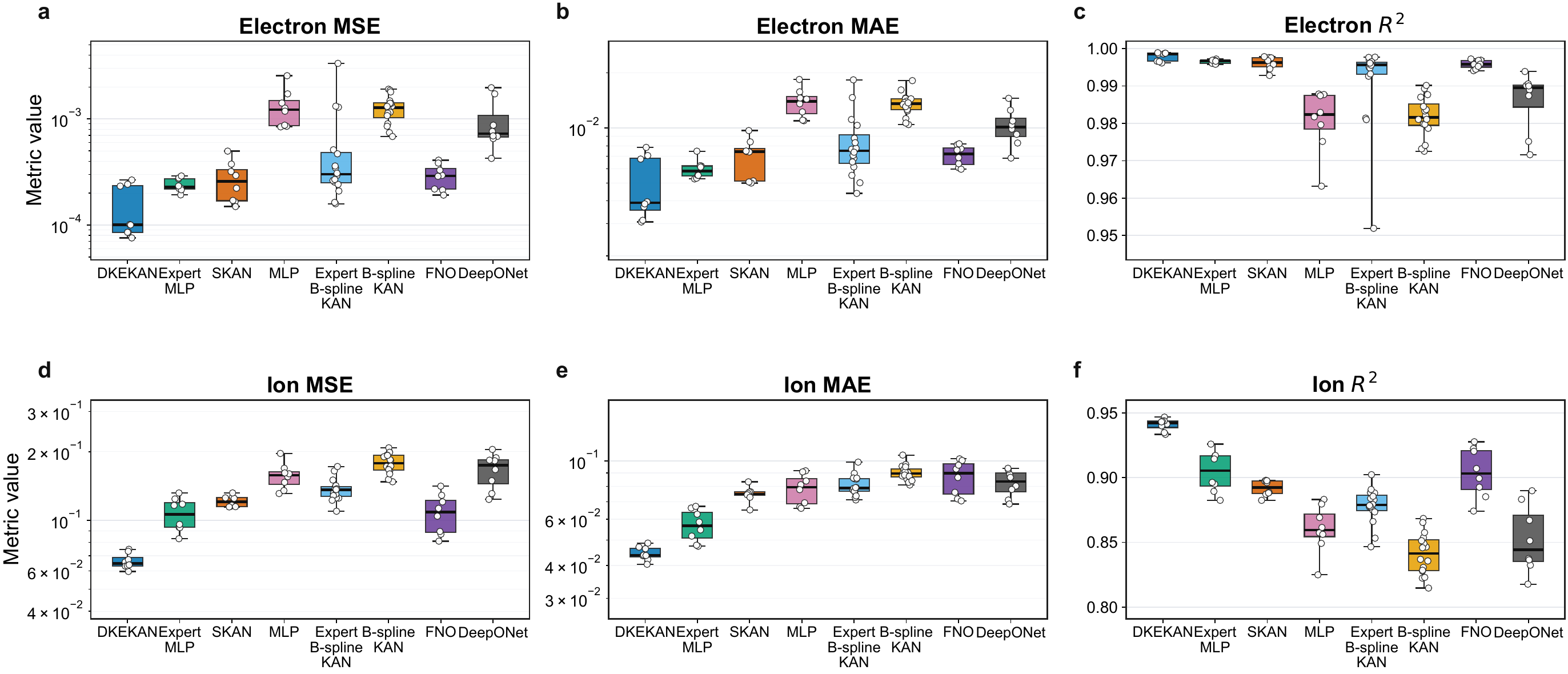}
    \caption{Test-set performance distribution over the model-specific hyperparameter space. Each box summarizes the formal full-test performance of all evaluated hyperparameter configurations for a given model. The boxes indicate the median and interquartile range, the whiskers span the minimum-to-maximum range, and the individual markers denote individual configurations.}
    \label{fig:hyperparameter_robustness}
\end{figure}

Fig.~\ref{fig:hyperparameter_robustness} further shows the test-set performance distribution over the model-specific hyperparameter search space. This figure is useful because the best result alone may not fully reflect the stability of a model. For DKEKAN, the performance remains favorable over the tested configurations, and the spread of the results is relatively controlled. In contrast, several baseline models show either larger variations or consistently higher error levels. This suggests that the advantage of DKEKAN is not only due to one specific hyperparameter choice, but is also reflected in the overall search-space behavior.

\begin{figure*}[!htbp]
    \centering
    \includegraphics[width=0.95\textwidth]{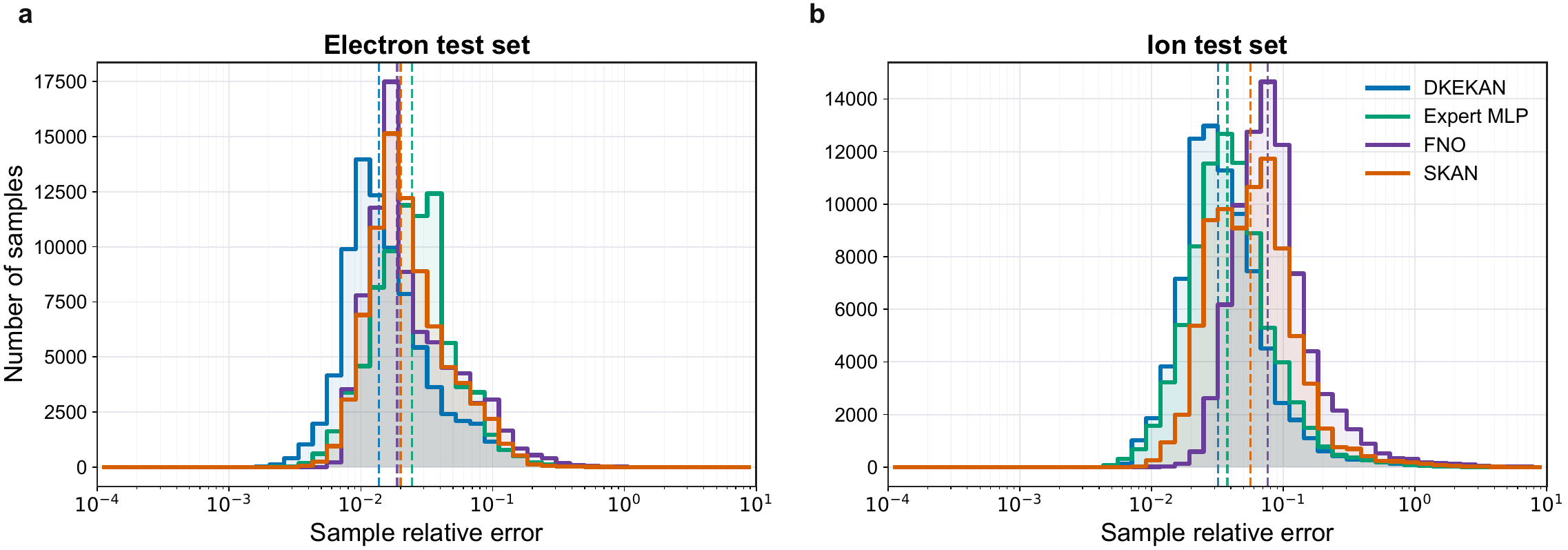}
    \caption{Test-set sample-wise relative-error distributions for the four leading surrogate models. The four models are selected according to their average rank over the formal full-test MSE, MAE, and $R^2$ across both ion and electron cases. Lower ranks correspond to better MSE and MAE, while higher $R^2$ receives a better rank. The selected models are DKEKAN, Expert MLP, FNO, and SKAN, with mean ranks of 1.00, 2.67, 3.33, and 4.83, respectively. Dashed vertical lines denote the median sample-wise relative error.}
    \label{fig:relative_error_distribution}
\end{figure*}

To examine the error behavior at the sample level, Fig.~\ref{fig:relative_error_distribution} compares the relative-error distributions of the four leading surrogate models. The relative error is defined as
\begin{equation}
    \varepsilon_F =
    \frac{\left|F-\hat{F}\right|}
    {\max \left|F\right|},
    \label{eq:relative_error}
\end{equation}
where $F$ denotes the reference solution, $\hat{F}$ denotes the model prediction, and $\max |F|$ is used for normalization.

Compared with the other leading models, the relative-error distribution of DKEKAN is more concentrated near zero. In other words, most of its predictions stay closer to the reference DKE solutions. This observation is in line with the quantitative metrics in Table~\ref{tab:formal_test_accuracy_runtime}, and it also shows that the improvement of DKEKAN is not only reflected in averaged indicators, but also in the sample-wise prediction behavior.

\begin{figure}[H]
    \centering
    \includegraphics[width=0.95\linewidth]{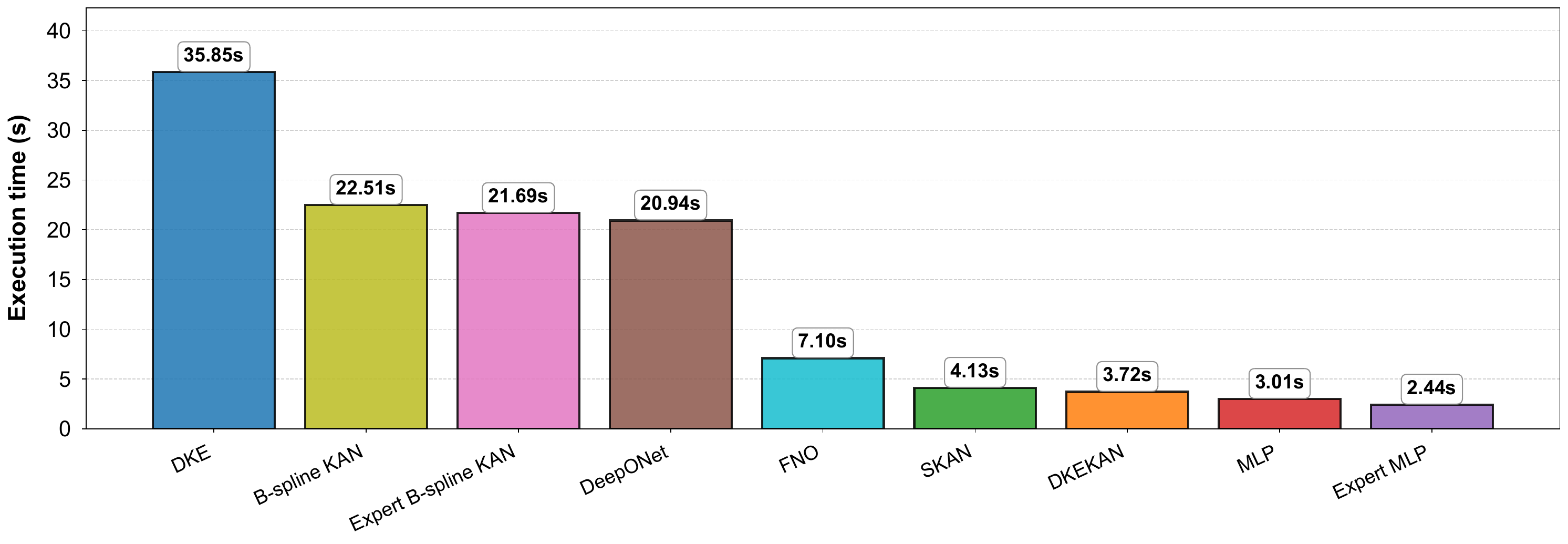}
    \caption{Execution time after model integration into the MATLAB-based physical solver. Compared with the original DKE solver, which requires 35.85 s for one complete evaluation, all surrogate models provide a clear reduction in end-to-end runtime.}
    \label{fig:runtime_comparison}
\end{figure}

The computational efficiency is then evaluated in the MATLAB-based NTVTOK workflow. Each model is tested by completing one NTV torque evaluation after being integrated into the physical solver. As shown in Fig.~\ref{fig:runtime_comparison} and Table~\ref{tab:formal_test_accuracy_runtime}, the original DKE solver requires 35.85 s for one complete evaluation. After replacing it with surrogate models, the runtimes of SKAN, Expert MLP, MLP, Expert B-spline KAN, B-spline KAN, DeepONet, FNO, and DKEKAN are 4.13, 2.44, 3.01, 21.69, 22.51, 20.94, 7.10, and 3.72 s, respectively.

Among all surrogate models, Expert MLP gives the shortest runtime. However, its prediction accuracy is lower than that of DKEKAN on both ion and electron test sets. DKEKAN is slightly slower than Expert MLP and MLP, but the additional cost is small compared with the improvement in accuracy. For NTV torque calculation, this trade-off is important, since the reliability of the final physical quantities depends strongly on the quality of the DKE solution.

It should also be kept in mind that the DKE solver is only one part of the full NTV torque calculation. The complete workflow also involves input preparation, post-processing, and other auxiliary computations. Therefore, the runtime difference between DKEKAN and the fastest lightweight neural models has a limited influence on the total computational cost. By contrast, the improvement in DKE prediction accuracy is more directly related to the reliability of the final simulation results.

Overall, DKEKAN reduces the runtime from 35.85 s to 3.72 s compared with the original DKE solver, while achieving the best prediction accuracy among the tested models. This corresponds to nearly one order of magnitude acceleration in the complete NTV torque evaluation. Considering the formal test accuracy, robustness over hyperparameter settings, sample-wise error distribution, and practical runtime after deployment, DKEKAN provides the most balanced performance among the compared models.

\subsection{Integration into NTVTOK and NTV prediction}
\label{subsec:ntvtok_integration}

\begin{figure}[!htbp]
    \centering
    \includegraphics[width=0.9\linewidth]{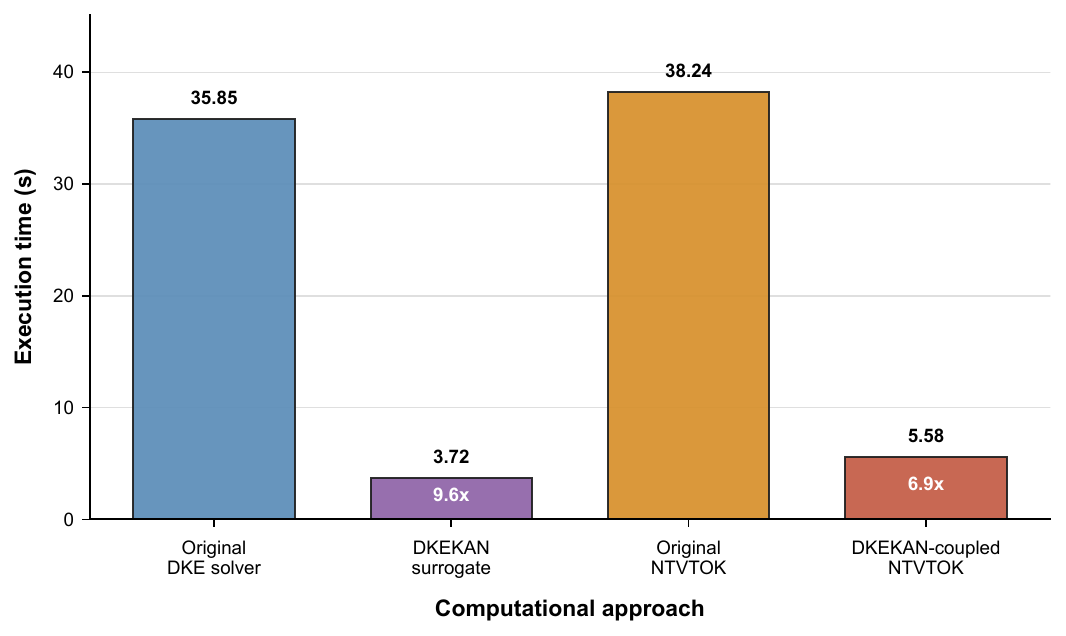}
    \caption{Computational efficiency of DKEKAN in standalone DKE solving and coupled NTVTOK simulations. The comparison reports the wall-clock runtime of the conventional NTVTOK workflow and the DKEKAN-accelerated workflow.}
    \label{fig:acceleration_efficiency}
\end{figure}

The DKEKAN model was integrated into the NTVTOK workflow using the electron and ion models with the hyperparameter configurations that achieved the best validation performance. The resulting system-level acceleration is shown in Fig.~\ref{fig:acceleration_efficiency}. Replacing the conventional DKE solver with DKEKAN leads to a marked reduction in computational cost for both standalone DKE evaluation and coupled NTVTOK simulations. The standalone DKE solving time is reduced from 35.85~s to 3.72~s, corresponding to a speedup of approximately \(9.6\times\). For the fully coupled NTVTOK workflow, the total runtime decreases from 38.24~s to 5.58~s, giving an overall acceleration of approximately \(6.9\times\). These results indicate that DKEKAN provides a practical route for accelerating engineering-scale NTV calculations while preserving the coupled simulation structure of NTVTOK.

For visualization, the NTV torque density is transformed using a signed logarithmic mapping~\cite{Yan2025NTVTOKML},
\begin{equation}
    T_{\mathrm{NTV}}^{\log}
    =
    \operatorname{sgn}\left(T_{\mathrm{NTV}}\right)
    \log_{10}
    \left(
    10^5 \left|T_{\mathrm{NTV}}\right| + 1
    \right).
    \label{eq:ntv_signed_log}
\end{equation}
This transformation preserves the sign of the torque while compressing its dynamic range, which is useful for comparing profiles that contain both weak and strong torque regions.

Because the NTV torque density can vary by several orders of magnitude along the radial coordinate, the mean relative error is used to quantify the global prediction accuracy. At a spatial location \(s\), the pointwise relative error is defined as
\begin{equation}
    \varepsilon_s
    =
    \frac{
    \left| T_{\mathrm{NTV},s} - \hat{T}_{\mathrm{NTV},s} \right|
    }{
    \max_s \left| T_{\mathrm{NTV},s} \right|
    },
    \label{eq:ntv_relative_error}
\end{equation}
where \(T_{\mathrm{NTV},s}\) and \(\hat{T}_{\mathrm{NTV},s}\) denote the reference and DKEKAN-predicted NTV torque densities, respectively. The reported mean relative error is obtained by averaging \(\varepsilon_s\) over all radial locations. This normalization avoids excessive weighting of local low-amplitude regions and gives a consistent measure of profile-level agreement.

\subsubsection{In-distribution NTV torque prediction}
\label{subsubsec:ntv_indistribution}

\begin{figure}[H]
    \centering
    \includegraphics[width=0.75\linewidth]{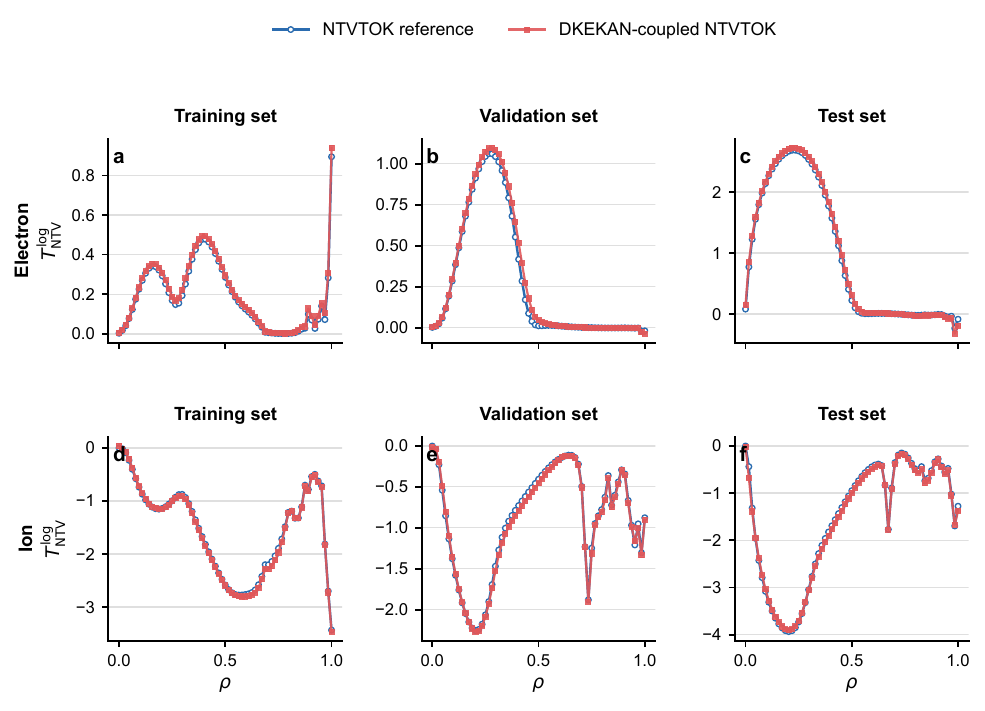}
    \caption{In-distribution NTV torque density prediction by the DKEKAN-coupled NTVTOK workflow. Representative plasma configurations from the training, validation, and test sets are compared against the reference NTVTOK results for both electron and ion torque density profiles.}
    \label{fig:ntv_indistribution_comparison}
\end{figure}

The in-distribution performance of DKEKAN was first evaluated on representative plasma configurations drawn from the training, validation, and test sets. As shown in Fig.~\ref{fig:ntv_indistribution_comparison}, the DKEKAN-coupled NTVTOK predictions reproduce the reference NTVTOK torque density profiles across all three data splits.

The electron torque predictions yield mean relative errors of 0.886\%, 2.255\%, and 1.922\% on the training, validation, and test configurations, respectively. The corresponding ion torque errors are 0.480\%, 0.685\%, and 0.828\%. The validation and test results remain at the same percent-level accuracy as the training results, indicating that the model does not merely memorize the training samples but retains stable predictive accuracy on unseen in-distribution cases. Overall, DKEKAN captures both the amplitude and the radial structure of the electron and ion NTV torque density profiles with errors at the percent level.

\subsubsection{Out-of-distribution generalization tests}
\label{subsubsec:ntv_ood}

To examine whether DKEKAN remains reliable outside the parameter range used for training, two out-of-distribution tests were considered. The first is a controlled profile extrapolation test, in which selected EAST plasma parameters, including the electron density \(N_e\), electron temperature \(T_e\), and ion temperature \(T_i\), are scaled by a factor of two. The second is a cross-device test based on ITER-like profiles. These two cases are designed to probe different forms of distribution shift: the controlled test isolates profile-amplitude extrapolation within an EAST-based setting, whereas the ITER case evaluates transfer to a substantially different device regime.

\begin{figure}[H]
    \centering
    \includegraphics[width=1\linewidth]{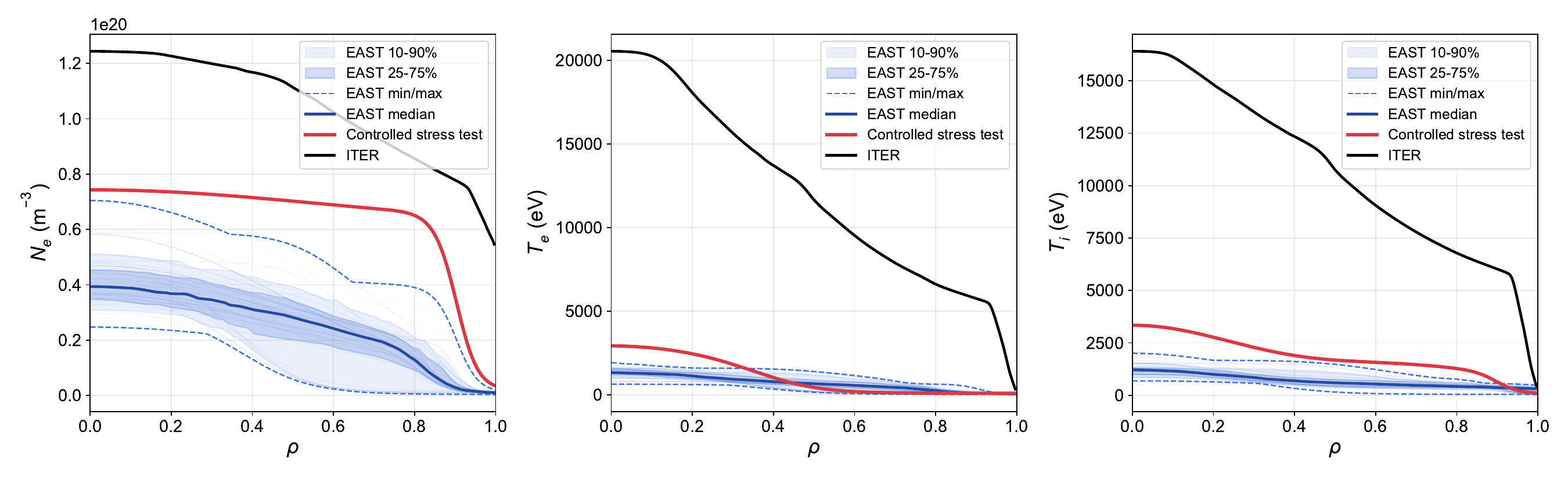}
    \caption{Radial distributions of the key plasma profiles used for the out-of-distribution tests. The blue shaded regions represent the EAST in-distribution database, with light and dark bands denoting the 10--90\% and 25--75\% percentile intervals, respectively. The dashed blue curves indicate the EAST minimum and maximum envelopes, and the solid blue curve gives the EAST median profile. The red curves correspond to the controlled \(2\times\) profile extrapolation test, while the black curves denote the ITER cross-device profiles. The three panels show \(N_e\), \(T_e\), and \(T_i\) as functions of the normalized radial coordinate \(\rho\).}
    \label{fig:ood_profile_distribution}
\end{figure}

The parameter distributions used in these tests are shown in Fig.~\ref{fig:ood_profile_distribution}. The controlled \(2\times\) profiles lie beyond the original EAST database by construction. The ITER profiles differ more strongly, particularly in the temperature channels, and therefore provide a more demanding assessment of cross-device generalization.

\begin{figure}[H]
    \centering
    \includegraphics[width=0.75\linewidth]{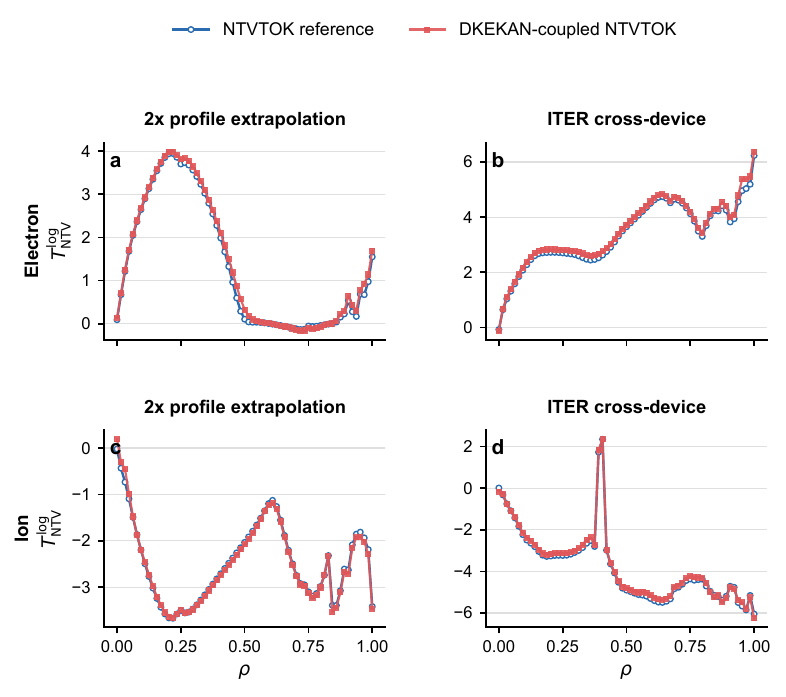}
    \caption{Out-of-distribution NTV torque density prediction using the DKEKAN-coupled NTVTOK workflow. The left column shows the controlled \(2\times\) profile extrapolation test, and the right column shows the ITER cross-device test. The upper and lower rows correspond to electron and ion NTV torque density profiles, respectively. Blue curves with circle markers denote the reference NTVTOK results, while red curves with square markers denote the DKEKAN-coupled NTVTOK predictions. The plotted quantity is the signed-logarithmic torque density \(T_{\mathrm{NTV}}^{\log}\).}
    \label{fig:ntv_ood_comparison}
\end{figure}

The corresponding NTV torque predictions are presented in Fig.~\ref{fig:ntv_ood_comparison}. In the controlled \(2\times\) profile extrapolation test, the mean relative errors are 1.797\% for the electron torque density and 1.409\% for the ion torque density. In the ITER cross-device test, the corresponding errors are 1.112\% and 2.438\%, respectively. Although the ITER profiles are far outside the EAST training distribution, the predicted torque profiles remain closely aligned with the NTVTOK references, including the main radial variations and the localized extrema in the signed-logarithmic representation.

These results show that DKEKAN retains percent-level accuracy under both controlled extrapolation and cross-device transfer. The performance in the ITER case is particularly relevant, since it suggests that the learned DKE surrogate can remain effective in plasma regimes that are not directly represented in the EAST training database. This supports the use of DKEKAN as an efficient surrogate module for NTV prediction in broader fusion-relevant operating scenarios.

\FloatBarrier

\section{Conclusion}
\label{sec:conclusion}

In summary, this paper presents DKEKAN, a surrogate model that integrates a modular expert layer with a lightweight SKAN backbone for solving the drift kinetic equation in NTV torque modeling. The results indicate that incorporating the physical grouping prior of DKE variables through a deterministic modular input interface is beneficial for representing heterogeneous inputs and contributes substantially to the improved accuracy of DKEKAN. Comparative results confirm that DKEKAN gives the best overall performance on the DKE-solving task among the tested MLP, KAN, and neural-operator baselines, while its inference speed enables significant acceleration in the coupled NTVTOK simulation workflow. When coupled back into NTVTOK, DKEKAN reproduces electron and ion NTV torque profiles with percent-level errors for in-distribution cases, and retains this level of agreement under controlled profile extrapolation and ITER-like cross-device tests. These results support the use of DKEKAN as an efficient DKE-solving module for fast NTV prediction without replacing the physics-based torque-evaluation procedure. Future work will refine the expert-layer design and extend the present generalization study to broader magnetic configurations, additional harmonic components, and higher-dimensional plasma-parameter spaces.

\section*{Acknowledgements}

This work was supported by the National Key R\&D Program of China under Grant No.~2024YFE03010000.

\section*{Data availability statement}

The data that support the findings of this study are not publicly available due to restrictions associated with the experimental EAST discharge data and the NTVTOK simulation workflow. The processed data and trained model parameters that support the results of this study are available from the corresponding author upon reasonable request.

\end{document}